\documentclass[a4paper,12pt]{article}

\usepackage{graphicx}
\usepackage{float}
\usepackage{amsmath}
\usepackage{amsthm}		                                  
\usepackage{amssymb}
\usepackage[utf8]{inputenc}

\newcommand{\eq}[1]{\begin{equation} #1 \end{equation}}
\newcommand{\al}[1]{\begin{equation} \begin{aligned} #1 \end{aligned} \end{equation}}

\usepackage{graphicx}
\usepackage{mathtools}
\usepackage{amssymb}
\usepackage{amsfonts}
\usepackage[footnotesize]{caption}
\usepackage[font=scriptsize]{subcaption}
\usepackage{color}
\usepackage{braket}
\usepackage{cite}
\usepackage{hyperref}
\usepackage{url}
\usepackage{multirow}
\usepackage{relsize}
\usepackage{fullpage}
\usepackage{makecell}
\usepackage{blkarray}
\usepackage{fullpage}

\begin{document}

\begin{titlepage}
\begin{center}
{\bf \Large Starobinsky-like inflation in no-scale supergravity Wess-Zumino model with Polonyi term } \\[12mm]
Miguel Crispim Rom\~{a}o$^{\star}$%
\footnote{E-mail: {\tt m.crispim-romao@soton.ac.uk}},
Stephen~F.~King$^{\star}$%
\footnote{E-mail: \texttt{king@soton.ac.uk}}
\\[-2mm]
\end{center}
\vspace*{0.50cm}
\centerline{$^{\star}$ \it
School of Physics and Astronomy, University of Southampton,}
\centerline{\it
SO17 1BJ Southampton, United Kingdom }
\vspace*{1.20cm}

\begin{abstract}
{\noindent
We propose a simple modification of the no-scale supergravity Wess-Zumino model of 
Starobinsky-like inflation to include a Polonyi term in the superpotential.
The purpose of this term is to provide an explicit mechanism for supersymmetry breaking at the end of inflation.
We show how successful inflation can be achieved for a gravitino mass satisfying the strict upper bound
$m_{3/2}< 10^3$ TeV, with favoured values $m_{3/2}\lesssim\mathcal{O}(1)$ TeV.
The model suggests that SUSY may be discovered in collider physics experiments such as the LHC or the FCC.
}
\end{abstract}
\end{titlepage}

\section{Introduction and Motivation}

Inflation \cite{Guth:1980zm,Linde:1981mu,Mukhanov:1981xt,Albrecht:1982wi,Linde:1983gd} provides a well motivated framework to address some of the remaining challenges facing the Cosmological Standard Model \cite{Linde:2007fr}, such as the 
flatness and horizon problems, as well as the absence of cosmological relics and the
origin of cosmological fluctuations. 
According to the paradigm of slow-roll inflation \cite{Linde1990,LythRiotto1999}, inflationary dynamics is described by one (or several) scalar field(s) called inflaton(s). If the potential for such scalar field is flat enough, then its dynamics will produce a source of negative pressure, which is required to sustain a phase of cosmological exponential acceleration.
Although it is not possible to test inflation in a model independent way, 
it may be said that current observational data \cite{Ade:2015lrj} generally supports the inflationary paradigm and 
constrains the dynamics of candidate inflationary models.

Over the past decades many proposals for inflationary models have been put forward \cite{MartinRingevalVennin2014}. Many of these theoretically viable models,
including the simplest chaotic models
based on polynomial potentials such as $\phi^2$ or $\phi^4$,
have recently been excluded by cosmic microwave background (CMB) data from the 
Planck satellite \cite{Ade:2015lrj}. The most recent data favours inflation models with purely Gaussian fluctuations,
a spectral index $n_s\approx 0.96 \pm  0.007$ and low tensor-to-scalar ratio $r< 0.08$.
Of the surviving classes of models, some are based on non-minimal gravity, such as the $R^2$ inflation model Starobinsky
\cite{R2}, although similar predictions arise in Higgs inflation \cite{HI} and related models \cite{others}, as well as low scale inflation models based on hybrid inflation \cite{Copeland:1994vg}.
However, many of the surviving models suffer from being sensitive to possible Ultra-Violet (UV) physics. If not protected by symmetry, generic corrections to the potential can provide an $\mathcal{O}(1)$ contribution to the slow-roll parameter $\eta$ spoiling the predictions required for a successful inflation. This is known as the $\eta$-problem \cite{Linde1990,LythRiotto1999}, and is a theoretical challenge facing any inflationary model.

Supersymmetry (SUSY) has been extensively used in inflationary models,
since such theories generally allow better control over the high energy dynamics of scalars
 \cite{Ellis:1982dg,Ellis:1982ed,Ellis:1982ws}.
 In particular, according to the Lyth bound \cite{Lyth:1984yz}, low values of the tensor-to-scalar ratio 
 $r$ imply that the scale of inflation should be less than the Planck scale, 
  and SUSY provides a mechanism for maintaining a hierarchy of scales without fine-tuning.
The first implementations of supersymmetric models of inflation considered only global SUSY. But since the dynamics of inflation occur at early stages of the history of the universe and is sensitive to UV scales,
it is necessary to also consider local SUSY or supergravity (SUGRA) (for recent SUGRA inflation models see for example \cite{YanagidaSUGRA,AntushcSUGRAinflation}). In SUGRA the so-called $F$-term scalar potential is sensitive to the shape of the K\"ahler potential, and this can lead to the re-emergence of the $\eta$-problem in the presence of quadratic contributions to the K\"ahler potential. One solution to this problem is no-scale SUGRA \cite{Ellis:1984bf}, where the K\"ahler potential takes a logarithmic form and circumvents the above problem. No-scale SUGRA models are so-called because the
scale at which SUSY is broken is undetermined in the first approximation, and the scale of the effective potential responsible for inflation can be naturally much smaller than the Planck scale, as required.
Alternatives to no-scale SUGRA have also been proposed, for example using a non-compact Heisenberg symmetry
\cite{BG} or a shift symmetry \cite{Y,Davis:2008fv,klor}.

Recently, it was shown that, within a Wess-Zumino framework, where the superpotential consists 
of quadratic and cubic terms only
\cite{Wess:1974tw, Croon:2013ana}, no-scale SUGRA can behave like a Starobinsky inflationary model \cite{Ellis:2013nxa,Ellis:2013xoa} (see also \cite{Ellis:2016ipm} for a realisation within an $SO(10)$ model, and for other No-Scale inflation realisations \cite{NoScaleInflation}). 
Such a model was shown to be the conformal equivalent of an $R+R^2$ model of gravity for a particular point
in parameter space, and at this point, Starobinsky inflation was shown to emerge.
However, in the above approach, the mechanism of SUSY breaking was left unspecified,
and formally the gravitino mass is zero in the Starobinsky limit.

In this paper we consider the above model of Ellis, Nanopoulos, Olive (ENO) \cite{Ellis:2013nxa,Ellis:2013xoa}, 
together with a linear Polonyi term to the superpotential. The purpose of adding this term is to provide
an explicit mechanism for breaking SUSY. If the Polonyi term is dropped, the model will reduce to 
the ENO model with Starobinsky inflation in a particular limit. Including a Polonyi term,
we shall show how one may successfully perturb away
from the Starobinsky limit of the ENO model, whilst maintaining successful inflation and at the same time generate a gravitino mass with a strict upper bound $m_{3/2}< 10^3$ TeV, with favoured values $m_{3/2}\lesssim\mathcal{O}(1)$ TeV.
The model suggests that SUSY may be discovered in collider physics experiments such as the LHC and the FCC.

This work is organised as follows. In section~\ref{noscale} we review the no-scale K\"ahler potential
and the general result for the supergravity potential. In section~\ref{WZP} we give the superpotential of our model
and present the inflationary potential and find its global minimum. 
In section~\ref{inflation} we discuss inflation and analyse the parameter space of the model, 
showing how this leads to an upper bound on the gravitino mass.
Section~\ref{conclusion} concludes the paper.

\section{The no-scale K\"ahler potential}
\label{noscale}

An important part of model is described by the K\"ahler potential alone, namely the kinetic term and a simple formula for the scalar potential can be found using only the K\"ahler potential.
The no-scale K\"ahler potential is given by
\eq{
	K = -3 M^2_{Pl} \ln \left(\frac{T+T^*}{M_{Pl}}- \frac{|Z|^2}{3 M_{Pl^2}}\right) \ ,
	\label{K}} 
	where $T$ is a modulus field and $Z$ is another field in the hidden sector which will be responsible for inflation
	and SUSY breaking.

The kinetic term depends on the second derivatives of the K\"ahler potential,
or K\"ahler metric given by,
\eq{
 K^j_i=\partial^2 K / \partial \phi^{*i} \partial \phi_j,\
(K^{-1})^k_i K^j_k= \delta^j_i \ .
\label{Kij} }
Using Eqs.\ref{K},\ref{Kij}, we find the kinetic term in the Lagrangian,
\eq{
	\mathcal{L}_{K}=\frac{3}{\left(\frac{T+T^*}{M_{Pl}}-\frac{|Z|^2}{3 M_{Pl}^2}\right)^2}(\partial_\mu Z^*, \partial_\mu T^*)\begin{pmatrix}
		\frac{T+T^*}{3 M_{Pl}} & - \frac{Z}{3 M_{Pl}} \\
		-\frac{Z^*}{3 M_{Pl}} & 1
	\end{pmatrix}
	\begin{pmatrix}
		\partial^\mu Z \\
		\partial^\mu T
	\end{pmatrix} \ .
	\label{matrix}}
	
In order to proceed we follow the procedure in ENO \cite{Ellis:2013nxa,Ellis:2013xoa}, where it is assumed that the $T$ modulus is fixed (or stabilised) by some other mechanism to be equal to a constant value,
so that its derivatives vanish. Then only the (1,1) element of the matrix in Eq.\ref{matrix} is non-zero,
where this element depends on the real vacuum expectation value of the $T$ modulus,
\eq{c = \langle T+T^*\rangle .}
The canonical normalisation of the remaining $Z$ kinetic term is somewhat non-trivial to achieve since it is multiplied 
by a function of $|Z|^2$.
However, the problem has been solved by ENO \cite{Ellis:2013nxa,Ellis:2013xoa} who change variables from
$Z$ to $\chi$ where,
\eq{
	Z =  \sqrt{3 c M_{Pl}} \tanh \left(\frac{\chi}{M_{Pl}\sqrt{3}}\right)
	\label{chi}}
in terms of which the kinetic term reads
\eq{
	\mathcal{L}_{K}= \text{sech}^2 \left[\frac{2 \text{Im}( \chi)}{M_{Pl }\sqrt{3}}\right]
	(\partial_\mu \chi^*)(\partial^\mu \chi) \ .
}
In the limit $\text{Im}(\chi) \to 0$ we get a canonically normalised kinetic term for $\text{Re}( \chi) $, which is identified as the inflaton in this limit.

Leaving every dimension explicit, the supergravity potential is given by
\eq{
V = K^j_iF_jF^{*i}-3 e^{K/M_{Pl}^2}\frac{|W|^2}{M_{Pl}^2} \ ,
}

	where the supergravity $F$ term is given by
\eq{
F_i = -  e^{K/2M_{Pl}^2}(K^{-1})^j_i(W_j^*+W^*\frac{K_j}{M_{Pl}^2}) \ .
}	
	

Using these results one finds, for any choice of superpotential $W$, the potential:
\eq{
	V = \frac{1}{\left(\frac{T+T^*}{M_{Pl}}-\frac{|Z|^2}{3 M_{Pl}^2}\right)^2}\left| \frac{\partial W}{\partial Z} \right|^2 
	\label{VF}	\ .
}

The last result manifests the main characteristic of a no-scale model that the potential is positive semi-definite.

Another important quantity is the gravitino mass $m_{3/2}$, whose square is given by
\eq{
    m^2_{3/2} =  \frac{K^j_iF_jF^{*i}}{3M^2_{Pl}}= e^{K/ M^2_{Pl}} \frac{|W|^2}{M^4_{Pl}}
 \label{m32}   
\ ,
}
    where the quantities are to be evaluated at the ground state vacuum, and the second equality above assumes
    that $V=0$ at the minimum.
    
    Both the potential and the gravitino mass depend crucially on the superpotential $W$ to which we now turn.


\section{Wess-Zumino-Polonyi model of Inflation}
\label{WZP}

The superpotential of our model consists of a Wess-Zumino model with quadratic and trilinear terms, of the kind considered in ENO \cite{Ellis:2013nxa,Ellis:2013xoa}, together with a new linear Polonyi term that we are adding in order to break SUSY:

\eq{
	W = M^2 Z + \frac{\mu}{2}Z^2 - \frac{\lambda}{3} Z^3
\label{W}	\ .
}
In the limit that $M=0$, the Polonyi term vanishes and the model reduces to the Wess-Zumino model considered 
by ENO.

Using $W$ in Eq.\ref{W}, the potential in Eq.\ref{VF} becomes,
\al{
	V = \frac{1}{\left( \frac{c}{M_{Pl}} - \frac{|Z|^2}{3 M_{Pl^2}} \right)^2} \left| M^2 + \mu Z - \lambda Z^2 \right|^2
\label{V}	
\ .
}

Since the potential is positive semi-definite, it is minimised for $V=0$.
The minimum of the potential is given by setting the numerator equal to zero,
leading to a quadratic equation for $Z$ with solution,
\eq{\label{eq:vevZ}
	Z = \frac{1}{2 \lambda} \left( \mu \pm \sqrt{\mu^2 + 4 M^2 \lambda} \right) \ .
}

Using the reparametrisation of the field as in Eq.\ref{chi},
\eq{
  Z = \sqrt{3 c M_{Pl}} \tanh \left(\frac{\chi}{\sqrt{3} M_{Pl}}\right) \ ,
}
and writing $\chi$ in terms of its real and imaginary parts
\eq{\chi = \frac{1}{\sqrt{2}}(x+i y)\ , }
the potential in Eq.\ref{V} becomes,

\eq{
 V= a\ \text{sec}^2 \left( \sqrt{\frac{2}{3}} \frac{y}{M_{Pl}}\right)\ \left| \text{cosh}\left( \frac{x+iy}{M_{Pl}\sqrt{6}}\right)  \right|^4\ \left| b + f \tanh \left( \frac{x+iy}{M_{Pl}\sqrt{6}}\right) - \tanh^2 \left( \frac{x+iy}{M_{Pl}\sqrt{6}}\right) \right|^2 \ ,
}

where
\al{
a &= |3 \lambda M^2_{Pl}|^2 \\
b &= \frac{M^2}{3c\lambda M_{Pl}} \\
f &= \frac{\mu}{\lambda \sqrt{3 c M_{Pl}}}\\
\chi &= \frac{1}{\sqrt{2}}(x+i y) \ .
}
For real parameters, we have checked that the potential is always minimised by $y=0$ for any value of $x$
in the range of interest for inflation. Moreover, we have checked that the steepness of the potential in the $y$ direction
always exceeds that in the $x$ direction in the range of interest.

Setting $y=0$, we then identify the inflaton as $x$, with the inflationary potential
\al{
 V=  a\ \left| \text{cosh}\left( \frac{x}{M_{Pl}\sqrt{6}}\right)  \right|^4 \
 \left| b + f \tanh \left( \frac{x}{M_{Pl}\sqrt{6}}\right) - \tanh^2 \left( \frac{x}{M_{Pl}\sqrt{6}}\right) \right|^2 \ .
}

The value of $x$ at the global minimum of the potential is
\eq{\label{eq:vevx}
x_0 = \sqrt{6} M_{Pl} \tanh^{-1} \left( \frac{1}{2} (f \pm \sqrt{4 b +f^2}) \right) \ ,
}
which can also be obtained from Eq.\ref{eq:vevZ}. Without loss of generality, for concreteness we will choose the lower sign, so that inflation will happen with $x$ rolling from $x_* > x_0$.


The ENO limit amounts to taking $M\to 0$, which means $b\to 0$. Without loss of generality, we take $f$ to be positive and take the lower sign in Eq.\ref{eq:vevx}. Then, in the limit $b\to 0$,
we find that $x_0 \to 0$, which also means $Z\to 0$ and hence $W\to 0 $ at the end of inflation.
Then, since $V=0$ at the global minimum, and the $F$ term is zero, SUSY is unbroken, 
and the ENO limit implies a massless gravitino according to Eq.\ref{m32}.

In the ENO limit, Starobinsky-like inflation is found in the further limit that
\eq{
\lambda = \frac{\mu}{\sqrt{3 c M_{Pl}}} \ ,
}
which in our case would account to take
\eq{f=1 \ .}

\begin{figure}[H]
	\centering
	\includegraphics[width=0.75\textwidth]{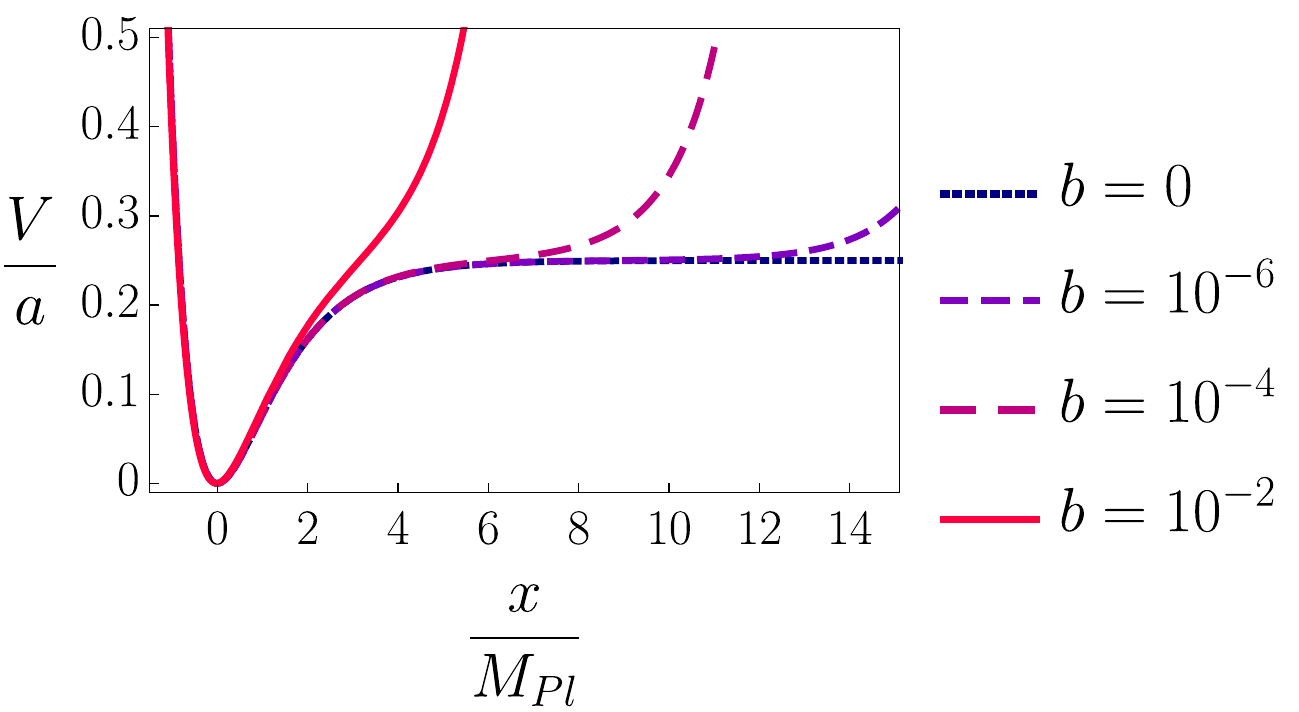}
	\caption{Comparison of the shape of the potential for different values of $b$,
	keeping $f=1$. As $b$ deviates from zero, 
	the value of the field at the global minimum $x_0$ shifts away from zero, while maintaining $V=0$.
	For small values of $b$ the potential retains a plateau where inflation will happen.	\label{fig:potential}}
\end{figure}

Therefore, Starobinsky-like inflation results from the limiting case of the above potential formula:
\al{
b&=0 \\
f&=1 \ .
}
In these limits, inflation occurs at $x_* \simeq 5.35 M_{Pl}$. As we can see in Figure \ref{fig:potential}, 
keeping $f=1$, a small value for $b$ represents a small deviation from the Starobinsky limit
of the ENO model. This means that we can expect a successful inflation scenario very similar to Starobinsky inflation for small values of $b$. This is of interest since, for non-zero $b$, SUSY is broken and the gravitino mass becomes non-zero
at the end of inflation.

\section{Inflation and parametric study of the model}
\label{inflation}

First we introduce the usual inflation functions. The potential slow-roll parameters are dimensionless and read
\eq{
	\epsilon = \frac{1}{2} M_{Pl}^2 \left( \frac{V^\prime}{V}\right)^2 
	}
and
\eq{
	\eta = M_{Pl}^2 \frac{V^{\prime\prime}}{V} \ ,
	}
where prime means derivative regarding the inflaton field, $x$.

From these we can derive two crucial dimensionless observables, the tensor-to-scalar ratio, $r$,
\eq{
	r \simeq 16 \epsilon
	}
and the scalar tilt, $n_s$,
\eq{
	n_s \simeq 1-6 \epsilon + 2 \eta \ .
	}

A last observable is the scalar amplitude, which reads\\
\eq{
	A_s = \frac{1}{24 \pi} \frac{V}{M_{Pl}^4 \epsilon}\label{eq:As} \ ,
	}
which is the only observable sensitive to the overall scale of the potential, i.e. of the parameter $a$. 

Furthermore, in order for inflation to be successful the dynamics of the inflaton has to significantly increase the scale factor. This is accounted by the N-folds quantity, that reads\\
\eq{N_* = \int_{x_*}^{x_f} \frac{1}{\sqrt{2\epsilon}}dx\ ,} 
where $x_*$ is the field value at which inflation starts, and $x_f$ the value at the end of inflation,
which we can safely approximate by $x_0$, the field value at the global minimum of the potential.

In the following we shall take the Starobinsky-like inflationary limit while allowing for a non-zero Polonyi term,
\al{
b &\neq 0 \\
f &= 1 \ .
}
In order for the potential to stabilise the scalar component of $Z$ with vanishing imaginary component, $y$, we enforce that Eq.\ref{eq:vevx} is real. In the limit $f=1$ this accounts for the constraint
\eq{
	-1 \leq \frac{1}{2} (1 \pm \sqrt{4 b +1}) \leq 1
}
with $4 b + 1 \geq 0$.
In their paper, ENO showed that deviations from $f=1$ limit  will change the shape of the potential and the range of the predictions of inflationary parameters without altering the massless nature of the gravitino. Therefore, we expect gravitino masses to be proportional to the value of $b$ and in henceforth keep the $f=1$ limit.

\begin{figure}[H]
	\centering
	\includegraphics[width=0.75\textwidth]{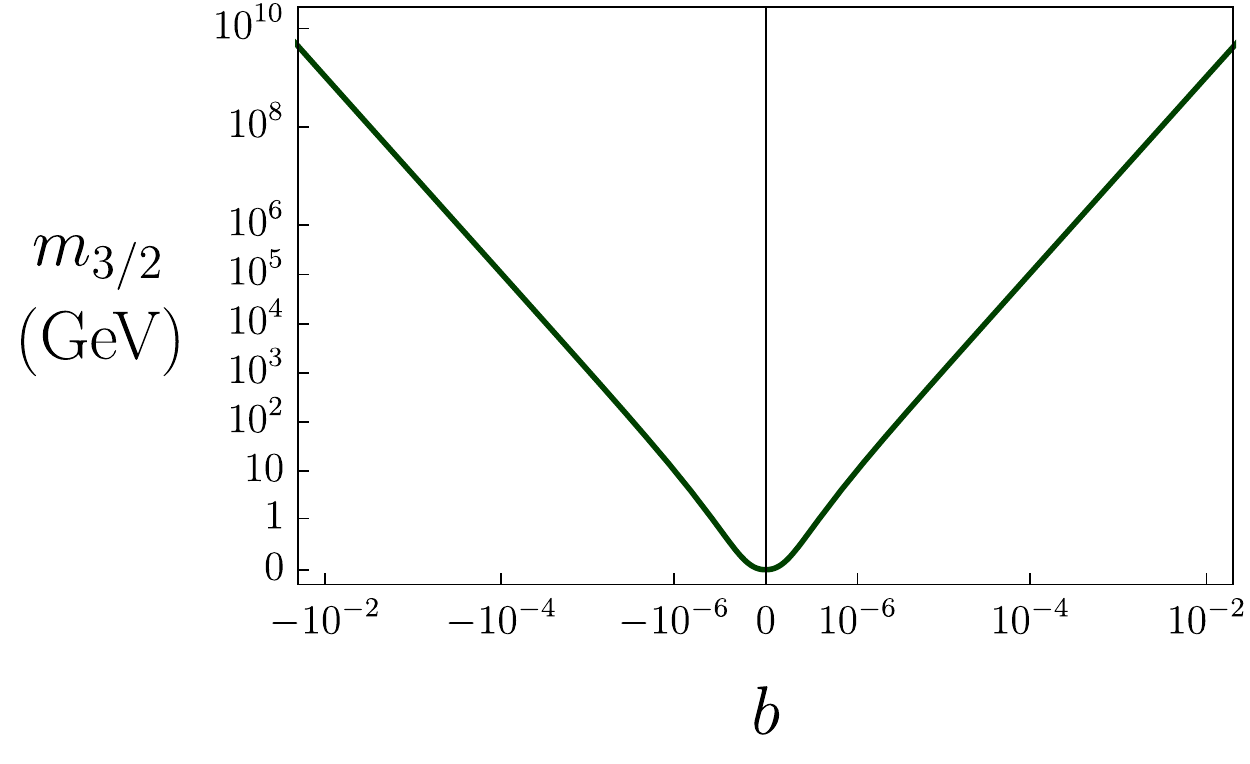}
	\caption{Dependency of $m_{3/2}$ on $b$, with fixed $x_*=5.35 M_{Pl}$.
		  \label{fig:m32b}}
\end{figure}

In order to study the inflationary dynamics, we need to specify the values for $(x_*,a,b)$ such that
\eq{
	50<N_*<60 \ .
	}

Since $N_* = N_*(x_*,b)$, the above conditions do not fix $a$. For that we use the central observational value for scalar spectrum amplitude, $A_s$. Therefore, we need only to scan a two-dimensional parameter space $(x_*,b)$ to study the inflationary dynamics of our model. 
In Figure~\ref{fig:m32b} we show the dependency of $m_{3/2}$ on $b$, with fixed $x_*=5.35 M_{Pl}$.
We have rescaled the results around the origin to show that for $b=0$, $m_{3/2}=0$, which is the ENO model limit.

\begin{figure}[H]
	\centering
	\includegraphics[width=0.75\textwidth]{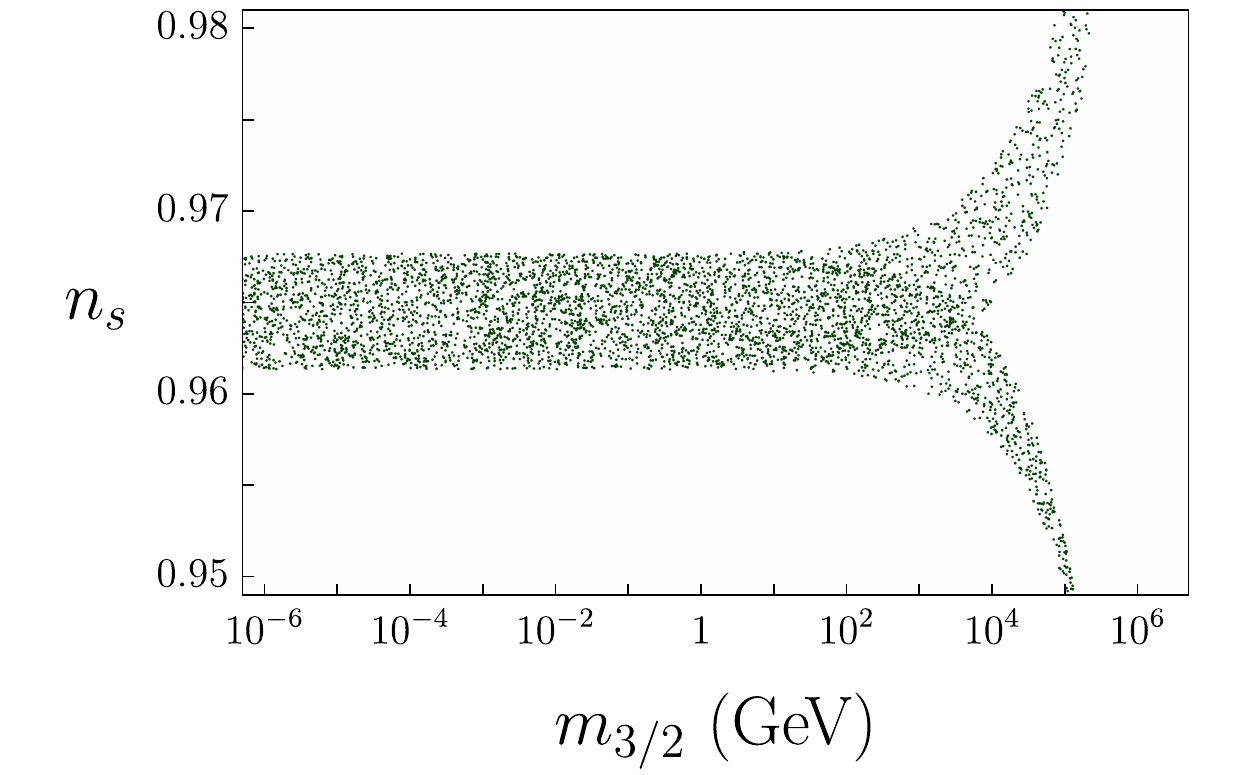}
	\includegraphics[width=0.75\textwidth]{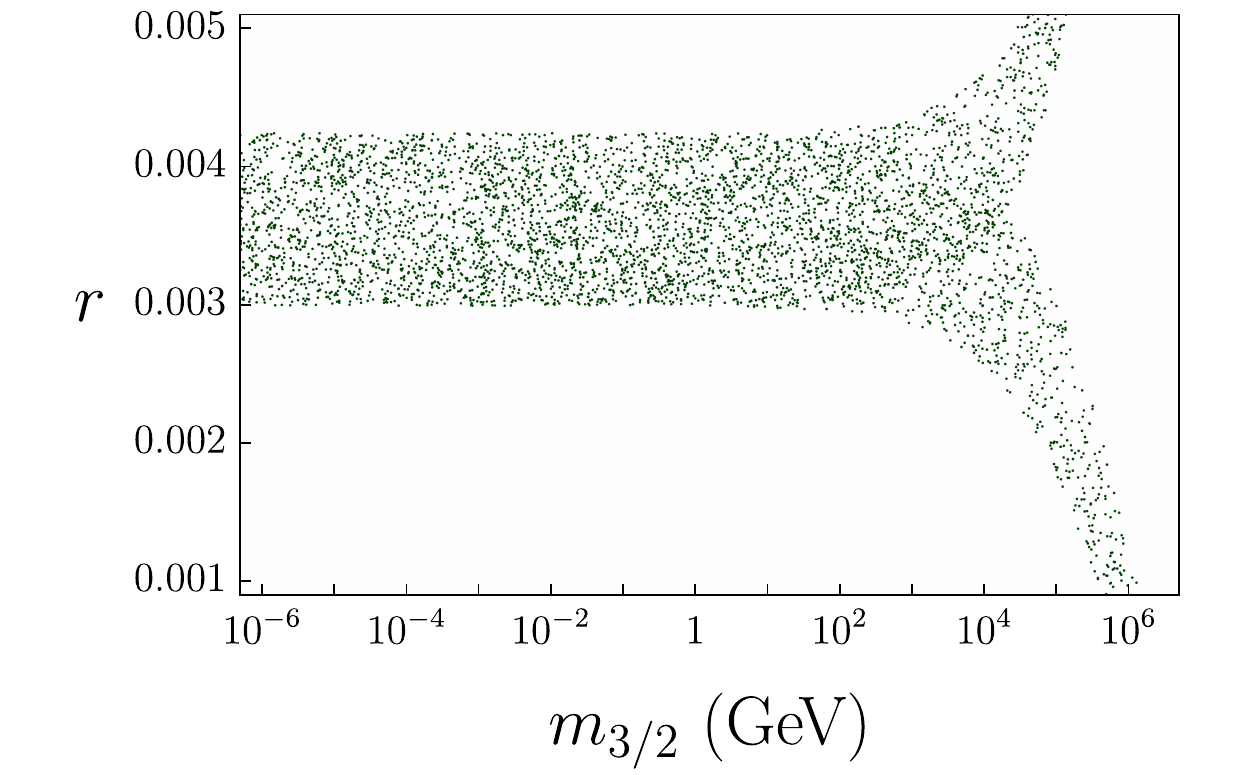}
	\caption{Scatter of points in the $(m_{3/2},n_s)$ and $(m_{3/2},r)$ planes,
	as a result of a scan over the parameter space $(x_*,b)$, keeping $f=1$.
	\label{fig:m32nsandr}}
\end{figure}

In Figure~\ref{fig:m32nsandr} we show the results of a scan over the parameter space $(x_*,b)$,
in the $(m_{3/2},n_s)$ and $(m_{3/2},r)$ planes. These results show that successful inflation can be achieved for a gravitino mass satisfying the strict upper bound
$m_{3/2}< 10^6$ GeV, with favoured values $m_{3/2}\lesssim\mathcal{O}(10^3)$ GeV.

\begin{figure}[H]
\centering
\includegraphics[width=0.75\textwidth]{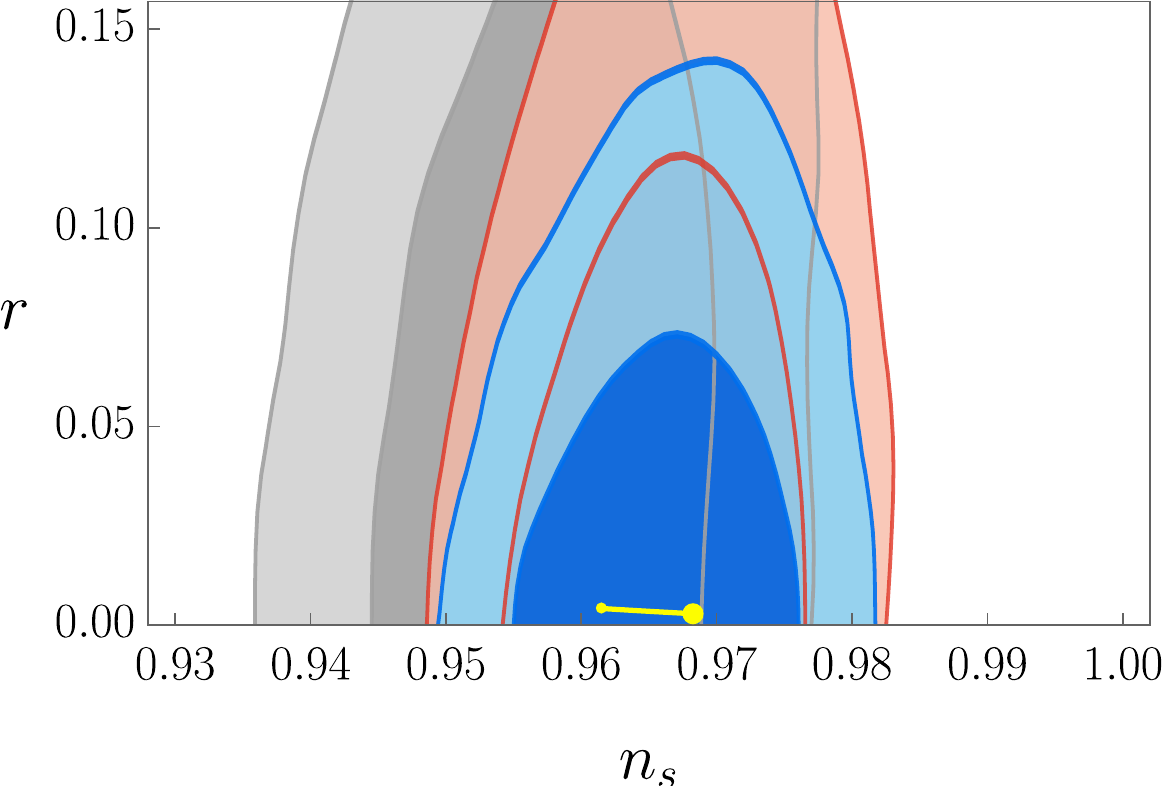}
\label{fig:nsr}
	\caption{$(n_s,r)$ predictions (in yellow) with the restriction $m_{3/2}\lesssim\mathcal{O}(10^3)$ GeV. The small endpoint 
	of the yellow bar represents $N_* = 50$, while large endpoint represents $N_*=60$. The predictions are compared to the PLANCK 2013 (Grey), PLANCK TT + lowP (Pink), and PLANCK TT, TE, EE + lowP (Blue) observational constraints \cite{Ade:2015lrj}\label{Planck}}
\end{figure}

In Figure~\ref{Planck} we show the subset of $(n_s,r)$ predictions, applying the restriction $m_{3/2}\lesssim\mathcal{O}(10^3)$ GeV, in the plane of the latest Planck results.

Further analysing the results, we found that the value of the inflaton field at the start of inflation satisfies the bounds
\eq{ 5.24 \lesssim \frac{x_*}{M_{Pl}} \lesssim 5.45 }
for the region where $m_{3/2} \lesssim \mathcal{O}(10^3)$ GeV.

Finally, in order to determine the value of the overall coefficient of the potential, $a$, we turn to the scalar amplitude Eq.\ref{eq:As}. This can be written as
\eq{
	A_s = a f(x_*,b) \ ,
}
where $f$ is some function of $x_*$ and $b$ and therefore, for each point $(x_*,b)$, we can determine $a$ by using the observed central value $10^9 A_s \simeq 2.2$. For the region of the parameter space yielding $m_{3/2}\lesssim \mathcal{O}(10^3)$ GeV, we find
\eq{
	1.27\times 10^{-10}\lesssim\frac{a}{M_{Pl}^4} \lesssim 1.81\times 10^{-10} \ .
}

The above results can be used to constrain the parameters of the K\"ahler and Superpotential. For example, the range for $a$ above means that $\lambda \simeq \mathcal{O}(10^{-5})$. Furthermore, if one assumes $c \simeq \mathcal{O}(M_{Pl})$, in the Starobinsky limit $f=1$, the bilinear mass term parameter becomes $\mu \simeq 10^{-5} M_{Pl}$. Under the same assumption, the results that suggest a light gravitino mass for $|b|\lesssim 10^{-5}$, and we find the Polonyi mass, $M$, to satisfy the bound
\eq{
	M \lesssim \mathcal{O}(10^{-5}) M_{Pl} \ .
}

\section{Conclusion}
\label{conclusion}

We have proposed a simple modification of the no-scale supergravity Wess-Zumino model of 
Starobinsky-like inflation to include a Polonyi term in the superpotential.
The purpose of this term is to provide an explicit mechanism for supersymmetry breaking at the end of inflation.
We have shown how successful inflation can be achieved for a gravitino mass satisfying the strict upper bound
$m_{3/2}< 10^3$ TeV, with favoured values $m_{3/2}\lesssim\mathcal{O}(10^3)$ GeV.
The model suggests that SUSY may be discovered in collider physics experiments such as the LHC or the FCC.

\begin{center}
{\bf Acknowledgments}
\end{center}
We would like Elena Accomando for valuable discussions. SFK acknowledges the STFC Consolidated Grant ST/L000296/1 and the European Union's Horizon 2020 Research and Innovation programme under Marie Sk\l{}odowska-Curie grant agreements 
Elusives ITN No.\ 674896 and InvisiblesPlus RISE No.\ 690575.
MCR acknowledges support from the FCT under
the grant SFRH/BD/84234/2012.

\end{document}